\documentclass{vgtc}

\onlineid{0}
\vgtccategory{Position Statement}
\vgtcinsertpkg

\graphicspath{{figures/}{./}}

\usepackage{times}

\usepackage{mathptmx}
\usepackage{booktabs}
\usepackage{balance}
\usepackage{tikz}
\usetikzlibrary{arrows.meta,positioning}

\title{When Direct Manipulation Becomes a Guess:\\
Productive Friction in AI-Mediated Multisensory Visualization}

\author{Anchit Mishra}
\affiliation{University of Waterloo}

\abstract{
Generative visualization increasingly embeds large language models within direct manipulation and multisensory interaction.
Speech, gaze, touch, gesture, sound, and haptics can make probabilistic inference feel like familiar, deterministic tool use.
I call this gap a \emph{deterministic-affordance mismatch}: deterministic interaction cues persist while AI weakens predictability, locality, reversibility, or provenance.
Reading the malleable interfaces of \emph{Iron Man 2} against systems from \emph{Blade Runner 2049}, I propose \emph{productive friction}, including cross-sensory renderings whose detail reflects model uncertainty.
Four frictions expose the inference boundary, match action to effect, make stochastic branches tangible, and attribute sensory agreement.
}


\begin{document}

\firstsection{Introduction}
\maketitle

In an \emph{Iron Man 2} scene, Tony Stark speaks, points, and expands a projected city model while JARVIS supplies the missing computation~\cite{Favreau2010}.
The artefacts persist, yet an intelligent agent interprets each incomplete action.
Current generative visualization resembles this fiction.
DynaVis turns language into persistent widgets~\cite{Vaithilingam2024}; Data Formulator delegates transformation to an AI agent~\cite{Wang2024}; and InterChat combines direct manipulation with LLM-based intent inference~\cite{Chen2025}.
They motivate a question: when does a probabilistic model adopt a deterministic tool's perceptual grammar?

Science-fiction interfaces act as diegetic prototypes shaping technological expectations~\cite{Kirby2010}.
They expose assumptions technical accounts leave implicit~\cite{Dourish2014}.
I identify this mismatch and four productive frictions that make consequential mediation perceptible.

\begin{figure*}[t]
  \centering
  \input{figures/scifi-comparison}
  \caption{Three science-fiction scenes expose different interface promises.
  (a) JARVIS interprets Stark's expansive gesture over a malleable model.
  (b) A backlit Denabase card keeps record comparison physically visible.
  (c) Duplicated hands in the Joi--Mariette merge disclose multiple embodied sources.
  Panels (a--b) are visualization cases; panel (c) is an ethical analogy.}
  \label{fig:scenes}
\end{figure*}

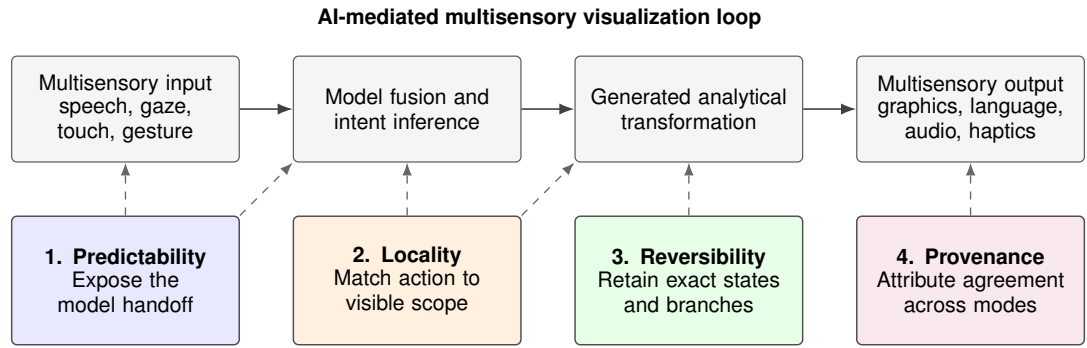
\begin{figure*}[t]
  \centering
  \begin{tikzpicture}[
  font=\sffamily\scriptsize,
  stage/.style={
    draw=black!65,
    rounded corners=2pt,
    minimum height=14mm,
    text width=28mm,
    align=center,
    fill=black!4,
    line width=0.45pt
  },
  gate/.style={
    draw=black!70,
    rounded corners=2pt,
    minimum height=17mm,
    text width=28mm,
    align=center,
    line width=0.55pt
  },
  flow/.style={-{Latex[length=2mm]}, line width=0.6pt, draw=black!75},
  link/.style={-{Latex[length=1.7mm]}, line width=0.5pt, draw=black!60, dashed}
]

\node[stage] (input) {Multisensory input\\[-1pt]\footnotesize speech, gaze, touch, gesture};
\node[stage, right=7mm of input] (fusion) {Model fusion and\\intent inference};
\node[stage, right=7mm of fusion] (transform) {Generated analytical\\transformation};
\node[stage, right=7mm of transform] (output) {Multisensory output\\[-1pt]\footnotesize graphics, language, audio, haptics};

\draw[flow] (input) -- (fusion);
\draw[flow] (fusion) -- (transform);
\draw[flow] (transform) -- (output);

\node[gate, fill=blue!9, below=7mm of input] (interpret) {\textbf{1. Predictability}\\[-1pt]\footnotesize Expose the model handoff};
\node[gate, fill=orange!12, below=7mm of fusion] (bound) {\textbf{2. Locality}\\[-1pt]\footnotesize Match action to visible scope};
\node[gate, fill=green!10, below=7mm of transform] (preserve) {\textbf{3. Reversibility}\\[-1pt]\footnotesize Retain exact states and branches};
\node[gate, fill=purple!9, below=7mm of output] (separate) {\textbf{4. Provenance}\\[-1pt]\footnotesize Attribute agreement across modes};

\draw[link] (interpret.north) -- (input.south);
\draw[link] (interpret.north east) -- (fusion.south west);
\draw[link] (bound.north) -- (fusion.south);
\draw[link] (bound.north east) -- (transform.south west);
\draw[link] (preserve.north) -- (transform.south);
\draw[link] (separate.north) -- (output.south);

\node[above=2.5mm of fusion, xshift=17.5mm, font=\sffamily\footnotesize\bfseries]
  {AI-mediated multisensory visualization loop};

\end{tikzpicture}
  \caption{An AI model can mediate the full multisensory visualization loop. Each friction answers a direct-manipulation promise weakened by model mediation: predictability, locality, reversibility, or provenance. At the output stage, sensory detail can also reflect model uncertainty.}
  \label{fig:pipeline}
\end{figure*}

\section{A Direct Action Is a Promise}

Direct manipulation rests on continuous representation, physical action, rapid feedback, and incremental, reversible operations~\cite{Shneiderman1983}.
Felt directness also depends on short semantic and articulatory distances between intention, action, and result~\cite{Hutchins1985}.
Multimodal systems complicate this contract.
They combine complementary signals, using one mode to resolve another's ambiguity~\cite{Oviatt1999}.
A gaze can identify an object while speech names an operation.

LLM mediation can weaken this contract at every stage in \cref{fig:pipeline}.
A small gesture may trigger global changes.
The same action may yield another result after context, sampling, or model updates.
An apparent undo may regenerate instead of restoring an exact state.
Fast feedback can hide these breaks because the action still \emph{feels} local, causal, and repeatable.

Multisensory output can strengthen that feeling.
People integrate visual and haptic estimates by apparent reliability~\cite{Ernst2002}; haptics can encode data through force, texture, or vibration~\cite{Paneels2010}.
A single model can coordinate graphics, narration, and haptic emphasis, making one error feel like several reinforcing sensations.
Exact-looking marks, tones, or haptic events can also overstate a coarse inference.
Designers can use purposefully low-fidelity sensory representations to communicate uncertainty in model output.
A coarse output could appear as a visual interval, a sonified band, or a diffuse haptic region.
Each replaces an exact mark, pitch, or detent.
The representation's perceptual resolution should match the inference's warranted resolution.
Exact source measurements should remain available for inspection.
No sense is inherently low fidelity; designers choose the detail encoded within it.
The mismatch therefore has four failure points: hidden inference weakens predictability; disproportionate effects weaken locality; regeneration weakens reversibility; and shared multisensory generation obscures provenance.

\section{Two Lessons from \emph{Blade Runner 2049}}

\emph{Blade Runner 2049} offers an operational counterpoint to JARVIS~\cite{Villeneuve2017}.
Its Denabase presents genomic and birth records through a mechanical, microfilm-like browser.
K advances and compares records until a duplicated pair appears.
In the morgue, dials and optical transitions expose increasing magnification from bone to microscopic serial number.
The design team describes both systems through physical, optical, microfiche, and mechanical references~\cite{Failes2017}.
Their friction makes evidence, comparison, and analytical labour visible.
Material controls provide no inherent truth, but these scenes keep intermediate states and the route from specimen to finding inspectable.

Joi's holographic artefacts disclose dependence on projectors and infrastructure.
Replicants possess material bodies, yet embodiment cannot settle their personhood.
The Joi--Mariette encounter distributes visual presence, touch, speech, and agency across several sources.
Productive friction should disclose provenance and AI mediation when aligned sources might otherwise merge.
It should never become a sensory test for deciding who, or what, deserves recognition.
The goal resembles Wheeler's ``reappearing tool'': mediation should become visible when transparency creates epistemic risk~\cite{Wheeler2019}.

\section{Four Frictions for Four Weakened Promises}

Each friction answers one failure point in the deterministic-affordance mismatch.
MacLean describes haptics as making digital transactions physical and confirmed~\cite{MacLean2008}; I extend this principle to mark the boundary between direct input and model inference.

\paragraph{1. Predictability---expose inference.}
JARVIS completes Stark's partial speech and gesture.
A real interface should expose that handoff before execution, consistent with human--AI guidance~\cite{Amershi2019}.
It can ghost the inferred target, operation, and cross-modal binding; a haptic detent, auditory transition, or visual threshold can separate exploration from commitment.

\paragraph{2. Locality---match action to effect.}
A small gesture should not silently trigger a global transformation.
Before execution, a scope halo should reveal every affected dataset, view, and encoding.
Effects exceeding the action's apparent scale require an explicit second commitment.
An inspectable trace exposes generated code, assumptions, and downstream changes.

\paragraph{3. Reversibility---make branches tangible.}
The Denabase and morgue sequences keep intermediate evidence visible as K advances.
Generative interfaces should likewise preserve exact checkpoints and expose alternative histories.
Undo restores the prior state; replay reuses captured inputs and settings; regenerate creates a labelled stochastic branch.
Persistent visual, auditory, or tangible tokens can support comparison between histories.

\paragraph{4. Provenance---attribute sensory agreement.}
The Joi--Mariette analogy keeps several embodied sources visible despite aligned outputs.
AI interfaces should likewise attribute agreement across graphics, narration, sonification, and haptic emphasis.
Coordinated cues need a common-source marker to avoid implying independent corroboration. At the same time,
measured data, external evidence, and model inference should remain separable.

\section{Conclusion}

Productive friction must remain selective.
Cognitive forcing can reduce overreliance but may be disliked~\cite{Bucinca2021}.
Designers should scale both friction and cross-sensory detail with uncertainty, consequence, and reversibility.
Every cue needs an accessible alternative; haptics should support access and embodiment without becoming mandatory evidence.

This paper is intended to provoke further discussion and debate rather than present an evaluated design framework;
it raises two empirical questions: 
\begin{enumerate}
    \item How should sensory detail scale with model uncertainty?
    \item How should friction scale with an operation's scope and reversibility?
\end{enumerate}
As visualization becomes more malleable and perceptually fluent, designers should dedicate effort to preserving directness (and thus, perceived user agency), expose AI limits, and prevent sensory detail from overstating model certainty. This way, interfaces of the future can make use of human sensory perception in a way that conveys AI mediation more realistically instead of exploiting multisensory perception to reinforce weak, uncertain, or inaccurate claims.

\section*{Figure Credits}

\Cref{fig:pipeline} was created by the author for this paper.
\Cref{fig:scenes} uses film frames for scholarly criticism and analysis.
Panel (a): \emph{Iron Man 2} (2010), \copyright~2010 MVL Film Finance LLC; source: Ortiz~\cite{Ortiz2020}.
Panels (b--c): \emph{Blade Runner 2049} (2017), \copyright~2017 Alcon Entertainment, LLC; sources: Territory Studio via Failes~\cite{Failes2017} and fxguide~\cite{Seymour2017}.
Composition and labels are by the author.

\balance
\bibliographystyle{abbrv-doi}
\bibliography{references}

\end{document}